\pdfoutput=1

\documentclass[aps,pre,twoside,twocolumn,floatfix,superscriptaddress,final]{revtex4-1}

\usepackage{mathptmx}
\usepackage{amsmath,amssymb}
\usepackage{microtype}

\usepackage{graphicx}

\usepackage[colorlinks,linkcolor=blue,citecolor=blue,urlcolor=blue]{hyperref}
\newcommand{\abs}[1]{\left|#1\right|}
\newcommand{\pdf}{\text{p.\hspace{1pt}d.\hspace{1pt}f.}}
\DeclareMathOperator{\var}{Var}

\usepackage{fancyhdr}
\fancypagestyle{first}{
   \fancyhf{}
   \fancyhead[C]{\large PHYSICAL REVIEW E \textbf{101}, 032102 (2020)}
   \fancyfoot[L]{\small 2470-0045/2020/101(3)/032102(8)}
   \fancyfoot[C]{\small 032102-\thepage}
}
\begin{document}

\title{Asymptotic behavior of the length of the longest increasing subsequences of random walks}

\author{J. Ricardo G. Mendon\c{c}a\href{https://orcid.org/0000-0002-5516-0568}{\includegraphics[trim=-5 0 0 0, scale=0.20]{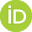}}}
\email{jricardo@usp.br}
\affiliation{Escola de Artes, Ci\^{e}ncias e Humanidades, Universidade de S\~{a}o Paulo, 03828-000 S\~{a}o Paulo, Brazil}
\affiliation{LPTMS, CNRS UMR 8626, Universit\'{e} Paris-Sud, Universit\'{e} Paris-Saclay, 91405 Orsay CEDEX, France}

\author{Hendrik Schawe\href{https://orcid.org/0000-0002-8197-1372}{\includegraphics[trim=-5 0 0 0, scale=0.20]{orcid_32x32}}}
\email{hendrik.schawe@u-cergy.fr}
\affiliation{Laboratoire de Physique Th\'{e}orique et Mod\'{e}lisation, CNRS UMR 8089, CY Cergy Paris Universit\'{e}, 95000 Cergy, France}
\affiliation{Institut f\"{u}r Physik, Universit\"{a}t Oldenburg, 26111 Oldenburg, Germany}

\author{Alexander K. Hartmann\href{https://orcid.org/0000-0001-6865-5474}{\includegraphics[trim=-5 0 -5 0, scale=0.20]{orcid_32x32}}}
\email{a.hartmann@uni-oldenburg.de}
\affiliation{Institut f\"{u}r Physik, Universit\"{a}t Oldenburg, 26111 Oldenburg, Germany}

\begin{abstract}
\makebox[5.25in][c]{\href{http://crossmark.crossref.org/dialog/?doi=10.1103/PhysRevE.101.032102&domain=pdf&date_stamp=2020-03-04}{\includegraphics[viewport=0 80 280 280, scale=0.045]{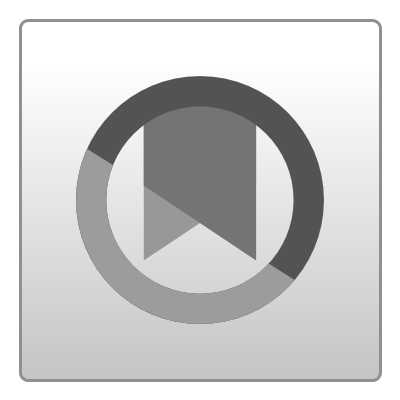}} (Sent 30 June 2019; revised manuscript sent 22 January 2020; accepted 13 February 2020; published 04 March 2020)} \\

We numerically estimate the leading asymptotic behavior of the length $L_{n}$ of the longest increasing subsequence of random walks with step increments following Student's $t$-distribution with parameter in the range $1/2 \leq \nu \leq 5$. We find that the expected value $\mathbb{E}(L_{n}) \sim n^{\theta}\ln{n}$ with $\theta$ decreasing from $\theta(\nu=1/2) \approx 0.70$ to $\theta(\nu \geq 5/2) \approx 0.50$. For random walks with distribution of step increments of finite variance ($\nu > 2$), this confirms previous observation of $\mathbb{E}(L_{n}) \sim \sqrt{n}\ln{n}$ to leading order. We note that this asymptotic behavior (including the subleading term) resembles that of the largest part of random integer partitions under the uniform measure and that, curiously, both random variables seem to follow Gumbel statistics. We also provide more refined estimates for the asymptotic behavior of $\mathbb{E}(L_{n})$ for random walks with step increments of finite variance. \\

{\noindent}DOI:~\href{https://doi.org/10.1103/PhysRevE.101.032102}{10.1103/PhysRevE.101.032102} 
\end{abstract}

\keywords{Correlated random variables, heavy-tailed random walk, time series, universality}

\maketitle \thispagestyle{first}


\section{\label{sec:intro}Introduction}

The longest increasing subsequence (LIS) problem is to find an increasing subsequence of maximum length of a given finite sequence of $n$ elements taken from a partially ordered set. Let $\mathcal{X}_{n}=(X_{1}, \dots, X_{n})$ be such a sequence---say, of real numbers. The longest (weakly) increasing subsequence of $\mathcal{X}_{n}$ is the longest subsequence $X_{i_{1}} \leq X_{i_{2}} \leq \cdots \leq X_{i_{L}}$ of $\mathcal{X}_{n}$ such that $1 \leq i_{1} < i_{2} < \cdots < i_{L} \leq n$, with $L$ the length of the LIS. There may be more than one ``longest'' increasing subsequence for a given $\mathcal{X}_{n}$, with different elements but of the same maximum length. Algorithmically, it takes $O(n)$ space and $O(n\log\log{n})$ time to find one LIS of a given sequence of $n$ elements \cite{sergei}.

Despite a superficial similitude, the LIS and the sequence of records of a time series---a rich traditional subject frequently studied in statistical mechanics---are unrelated and should not be confused. The LIS is the maximum-length increasing subsequence while the sequence of records is the subsequence of increasing maxima. For example, in the sequence $(2,1,6,4,3,5)$ the sequence of records is $(2,6)$, while the LIS are $(1,3,5)$, $(1,4,5)$, $(2,3,5)$, and $(2,4,5)$. Note also that the LIS of a time series is a much more intricate quantity than its set of records, since it depends on the entire series, not just on past events; it is a global property of the series. Algorithmically, the sequence of records can be obtained by one linear sweep through the sequence and can also be determined online (on the go) as the series progress, while more elaborate approaches are necessary to determine the LIS \cite{sergei}. As far as we currently understand, results on records are of no avail in the treatment of the LIS problem. The reader interested in the modern theory of records should consult \cite{nevzorov,godreche}. Examples of the application of the LIS can be found in computer science, where the sortedness of a data stream or list of items can be used, e.\,g., to decide whether the data have to be sorted again \cite{gopalan2007}, and also in the design of privacy preserving algorithms for publicly shared information through sequential data streams~\cite{bonomi2016}.

Initially, the LIS problem was considered in the mathematical literature for random permutations. The problem seems to have been first considered by Ulam in the early 1960s \cite{ulam}. The resolution of the LIS problem for random permutations culminated with the exact characterization of $L_{n}$ as a random variable distributed like $L_{n} \sim 2\sqrt{n}+\sqrt[6]{n}\,\chi$ with $\mathbb{P}(\chi \leq s) = F_{2}(s)$, the Tracy-Widom distribution for the fluctuations of the largest eigenvalue of a Gaussian unitary random matrix ensemble about its expected value \cite{t-widom,bdj99}. Comprehensive expositions on the LIS problem for random permutations appear in \cite{patience,romik}.

\begin{figure*}[t]
\centering
\includegraphics[viewport=30 25 550 305, scale=0.475, clip]{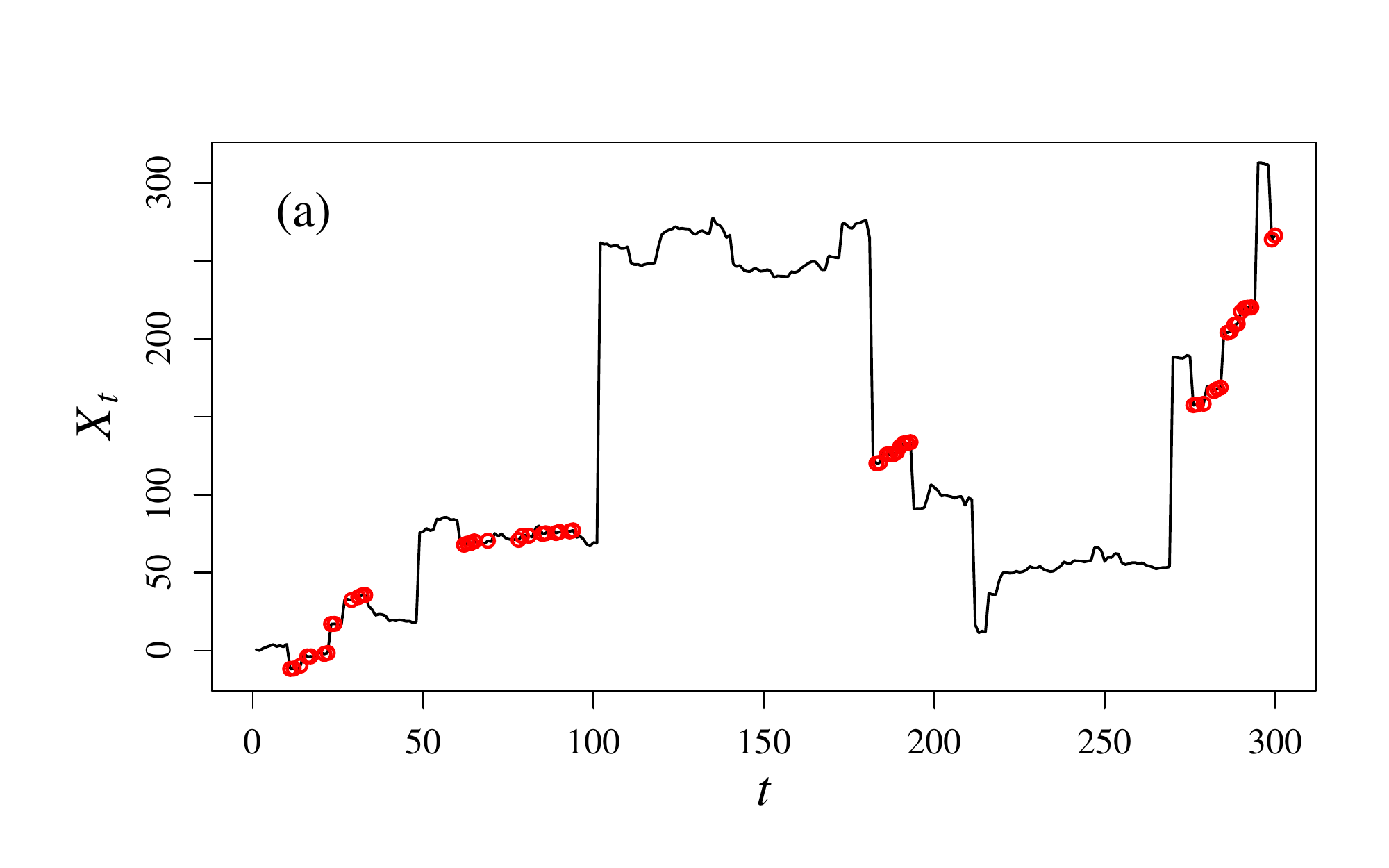} \hfill
\includegraphics[viewport=30 25 550 305, scale=0.475, clip]{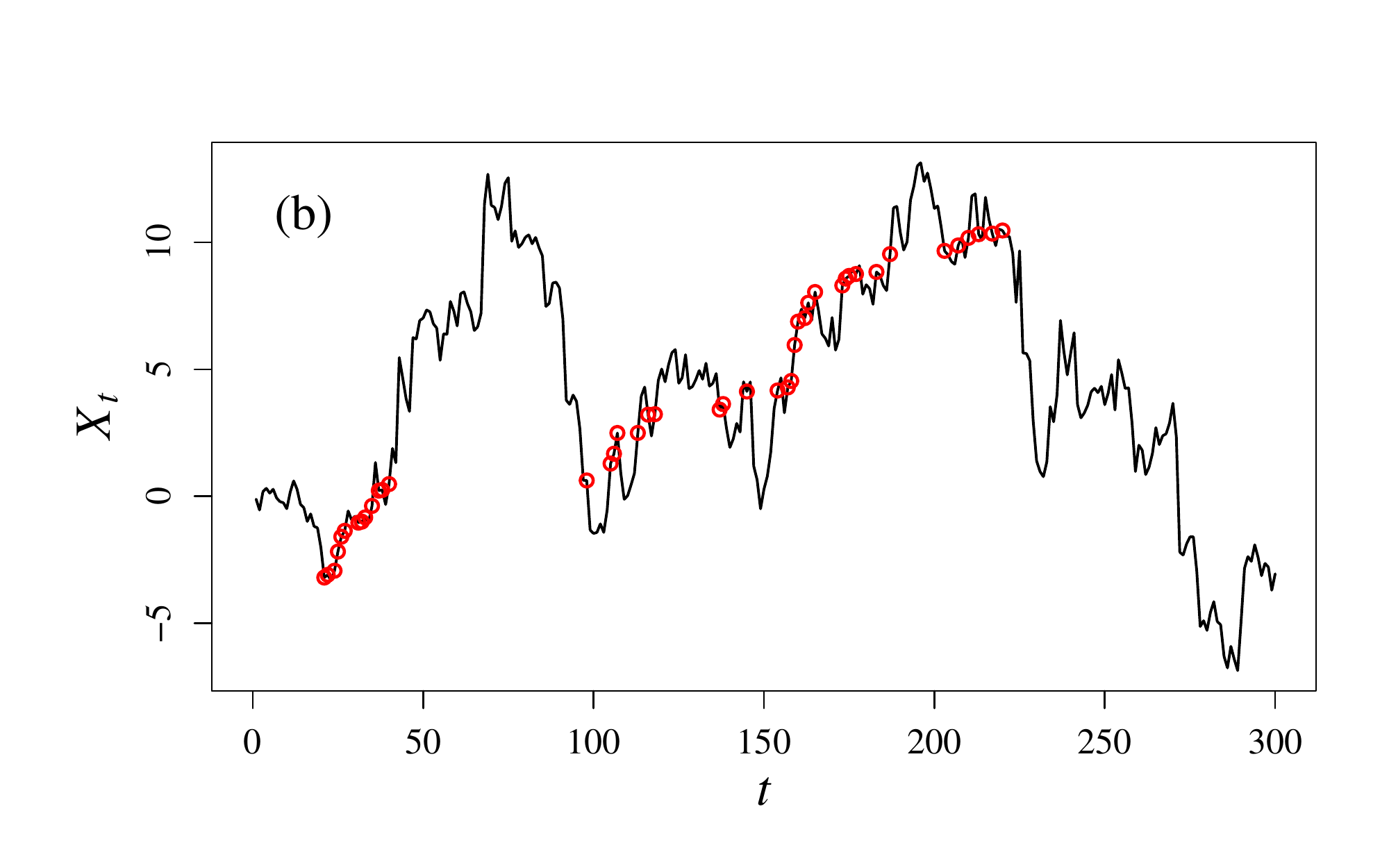}
\caption{Student's $t$ random walks of $300$ steps with parameters (a)~$\nu=1$ and (b)~$\nu=4$ together with one of their LIS each (small circles). Note the different vertical scales.}
\label{fig:rwrw}
\end{figure*}

Recently, another version of the LIS problem was posed \cite{angel,pemantle}: What is the behavior of the LIS of a random walk? Let $\mathcal{X}_{n}=(X_{1}, \dots, X_{n})$ be the sequence of terms of a random walk given by
\begin{equation}
\label{eq:rw}
X_{0}=0, \quad X_{t}=X_{t-1}+\xi_{t}, \quad t=1, \dots, n,
\end{equation}
with the $\xi_{t}$, $t=1, \dots, n$, independent random variables identically distributed according to some zero-mean symmetric probability distribution $\phi(\xi)$. The sequence $\mathcal{X}_{n}$~constitutes a time series of correlated random variables: if the expectation $\mathbb{E}(\xi^{2}) \neq 0$, then $\mathbb{E}(X_{t}X_{s}) \neq 0$. In \cite{angel}, the authors showed that when $\phi(\xi)$ has finite positive variance, then for all $\epsilon > 0$ and large enough $n$ the length $L_{n}$ of the LIS of $\mathcal{X}_{n}$ observes
\begin{equation}
\label{eq:finite}
c\sqrt{n} \leq \mathbb{E}(L_{n}) \leq n^{1/2+\epsilon}
\end{equation}
for some positive constant $c$. The upper bound in Eq.~(\ref{eq:finite}) does not rule out a logarithmic term and can actually be read as
\begin{equation}
\label{eq:log}
\mathbb{E}(L_{n}) \leq \sqrt{n}(\ln{n})^{a}
\end{equation}
for some $a \geq 0$. In~\cite{pemantle}, the authors further proved that the expected length of the LIS of a particular random walk with step lengths of ultra-heavy distribution without any finite (integer or fractional) moment scales with the length of the walk as
\begin{equation}
\label{eq:heavy}
n^{\beta_{0}-o(1)} \leq \mathbb{E}(L_{n}) \leq n^{\beta_{1}+o(1)},
\end{equation}
with non-sharp $\beta_{0} \simeq 0.690$ and $\beta_{1} \simeq 0.815$. Besides the bounds (\ref{eq:finite})--(\ref{eq:heavy}), little is known rigorously about the LIS of random walks.

Figure~\ref{fig:rwrw} displays two random walks of $300$ steps distributed according to a Student's $t$-distribution (see Sec.~\ref{sec:heavy}), one with parameter $\nu=1$ (the same as the Cauchy distribution) and the other with parameter $\nu=4$, together with one of their LIS each.

In order to improve our knowledge about the LIS of random walks, we ran Monte Carlo simulations \cite{newman1999,bigguide} to estimate the scaling of $L_{n}$ for several different distributions of step lengths \cite{lisrw,hartmann}. The simulations showed that
\begin{equation}
\label{eq:exp}
\mathbb{E}(L_{n}) \sim n^{\theta}
\end{equation}
with a non-universal scaling exponent $0.60 \lesssim \theta \lesssim
0.69$ for the heavy-tailed distributions of step lengths examined, with $\theta$ increasing as the distribution of step lengths becomes more heavy-tailed. For distributions of finite variance, one of us conjectured \cite{lisrw} that the asymptotic behavior of $\mathbb{E}(L_{n})$ in these cases is given by
\begin{equation}
\label{eq:conj}
\mathbb{E}(L_{n}) \sim \frac{1}{e}\sqrt{n}\ln{n} + \frac{1}{2}\sqrt{n}
\end{equation}
plus lower order terms, although the constants are presumed based on least squares adjustements. Moreover, it has been found that the empirical distribution of $L_{n}$ seems to be of the form
\begin{equation}
\label{eq:scaling}
f(L_{n}) = \frac{1}{\mathbb{E}(L_{n})}g\Big(\frac{L_{n}}{\mathbb{E}(L_{n})}\Big),
\end{equation}
with $\mathbb{E}(L_{n})$ given by Eq.~(\ref{eq:exp}) or (\ref{eq:conj}) depending on whether $\phi(\xi)$ has a finite second moment or not. Accordingly, when the step lengths are of finite vari\-ance $g(z)$ should be universal. Plots of $g(z)$ for different distributions of step lengths appear in Ref.~\cite{lisrw} (see also Fig.~\ref{fig:collapse}). The form (\ref{eq:scaling}) has been further tested in Ref.~\cite{hartmann}, that probed the distribution of $L_{n}$ into regions of very small probabilities for random walks with uniform increments $\xi \sim U(-1,1)$. The authors found very good agreement between Eqs.~(\ref{eq:conj}) and (\ref{eq:scaling}) and their data. They also estimated that the large deviation rate function $\Phi(L)$ associated with the distribution of $L_{n}$ by
\begin{equation}
\label{eq:rate}
f(L_n) \asymp \exp[-n\Phi(L_n)]
\end{equation}
behaves asymptotically, in the limit of large $n \to \infty$, like $\Phi(L) \sim L^{-1.6}$ in the left tail and $\Phi(L) \sim L^{2.9}$ in the right tail. Despite this characterization, the distribution $g(z)$ remains unknown. It is tempting to conjecture that the actual exponents in $\Phi(L)$ are, respectively, $L^{-3/2}$ and $L^{3}$, in which case they would be the same as those of the large deviation rate function of the Tracy-Widom $F_{2}$ distribution with the sides (left $\leftrightarrow$ right) and signs flipped \cite{satya}. Note, however, that unlike in Eq.~(\ref{eq:rate}), the large deviation rate function for the Tracy-Widom distribution in the right tail is defined by a relationship of the form $\exp[-\sqrt{n}\Phi(L)]$, i.\,e., with an unusual $\sqrt{n}$ scaling.

Interestingly, while the LIS of random permutations played a fundamental role in the development of a wealth of mathematical physics in the last few decades, the LIS of random walks did not receive any attention until very recently. Since the LIS finds applications in the analysis of data streams and time series, it is necessary to understand its fundamental behavior. In this paper we provide an updated account on the LIS problem for random walks. While in previous studies of the problem only specific distributions $\phi(\xi)$ of step increments were considered, in Section~\ref{sec:heavy} we employ a parametrized distribution, namely, the Student's $t$-distribution, considerably enlarging the class of random walks considered. This approach allows us to investigate the dependence of the scaling exponent $\theta$ in (\ref{eq:exp}) with the heavy tail index of $\phi(\xi)$. In Sec.~\ref{sec:remarks}, we further verify the proposed scaling form Eqs.~(\ref{eq:conj}) and (\ref{eq:scaling}) with independent data, this time taking the full distribution of the data into account. We also set down some remarks on the resemblance between the statistics of the LIS problem for random walks of finite variance and the random partition problem under the uniform measure. Section~\ref{sec:summary} concludes the paper and provides perspectives for further study.


\section{\label{sec:heavy}LIS of heavy-tailed random walks}

We investigate the behavior of the scaling exponent appearing in the relation $L_{n} \sim n^{\theta}$ for random walks with heavy-tailed distribution of step increments as a function of their characteristic index $\alpha$, defined by
\begin{equation}
\label{eq:tail}
\phi_{\alpha}(\abs{\xi} \gg 1) \sim \abs{\xi}^{-1-\alpha}.
\end{equation}
We want to check whether there exists a well defined relationship between $\theta$ and $\alpha$. In order to access a range of values of $\alpha \leq 2$ (such that $\mathbb{E}(\xi^{2})=\infty$), we employ Student's $t$-distribution \cite{student}
\begin{equation}
\label{eq:student}
\phi_{\nu}(\xi) = \frac{\Gamma\big[\frac{1}{2}(\nu+1)\big]}{\sqrt{\nu\pi}\,\Gamma\big(\frac{1}{2}\nu\big)} \bigg(1+\frac{\xi^2}{\nu}\bigg)^{\!-(\nu+1)/2},
\end{equation}
where $\Gamma(z)$ is the usual gamma function and $\nu$ is a parameter. This distribution appears in inference problems about unknown parameters (mean or variance or both) of a normal population. In statistical applications $\nu \geq 1$ is a natural number, but for modeling purposes $\nu$ can be taken a real positive number. When $\nu < \infty$, Student's $t$-distribution displays a heavy tail $\phi_{\nu}(\abs{\xi} \gg 1) \sim \abs{\xi}^{-1-\nu}$, with infinite variance if $\nu \leq 2$ and finite variance $\nu/(\nu-2)$ for $\nu > 2$. We see that $\nu$ plays the role of the tail index $\alpha$ in Eq.~(\ref{eq:tail}). Student's $t$-distribution becomes the Gaussian distribution in the limit $\nu \to \infty$.

For each parameter $\nu$ and walk length $n$, we generate $10^{4}$ realizations of $\mathcal{X}_{n}$, compute their $L_{n}$ and estimate the empirical average $\langle{L_{n}}\rangle$ and variance $\langle{L_{n}^{2}}\rangle-\langle{L_{n}}\rangle^{2}$. In our simulations $1/2 \leq \nu \leq 5$ and $10^{4} \leq n \leq 10^{7}$ (and up to $2\times 10^8$ for some values around $\nu=2$). Student's $t$ random deviates can be efficiently and reliably generated by the polar method \cite{bailey}. Whenever $\mathbb{E}(\xi^{2})$ is finite (i.\,e., $\nu>2$), we use normalized random variables $\xi/\!\sqrt{\mathbb{E}(\xi^{2})}$ for the step increments. To empirically obtain $\theta$ as a function of $\nu$ for all values of $\nu$, and led by the previous results embodied in Eqs.~(\ref{eq:exp}) and (\ref{eq:conj}), we fitted the data of the empirical averages $\langle L_n \rangle$ for all values of $\nu$ to
\begin{subequations}
\begin{equation}
\label{eq:theta}
\langle L_{n} \rangle \sim n^{\theta}(1+c\ln{n})
\end{equation}
and
\begin{equation}
\label{eq:gamma}
\langle{L_{n}^{2}}\rangle-\langle{L_{n}}\rangle^{2} \sim n^{\gamma}(1+d\ln{n})
\end{equation}
\end{subequations}
\begin{figure}[t]
\centering
\includegraphics[viewport=10 10 480 460, scale=0.245, clip]{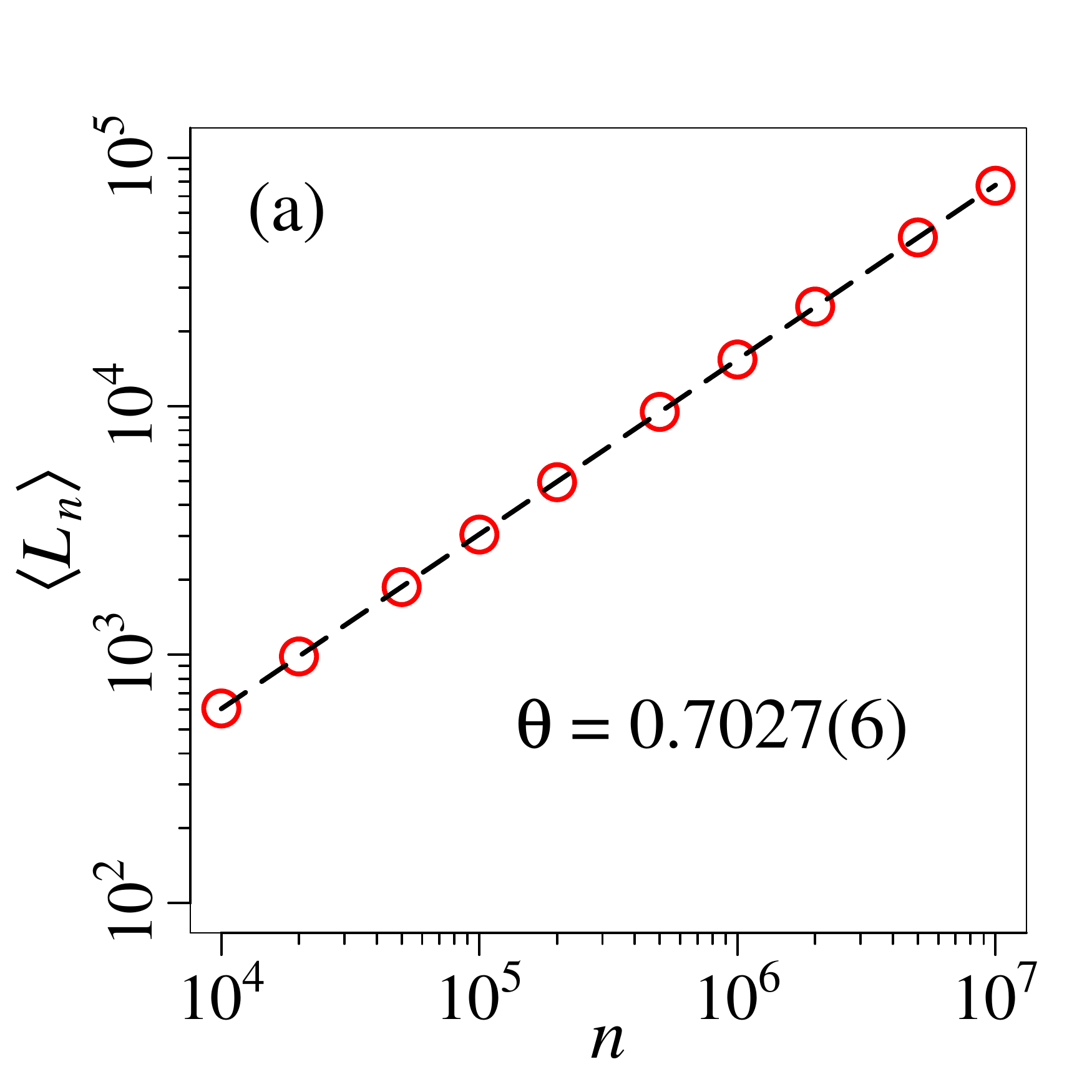} \hfill
\includegraphics[viewport= 0 10 480 460, scale=0.245, clip]{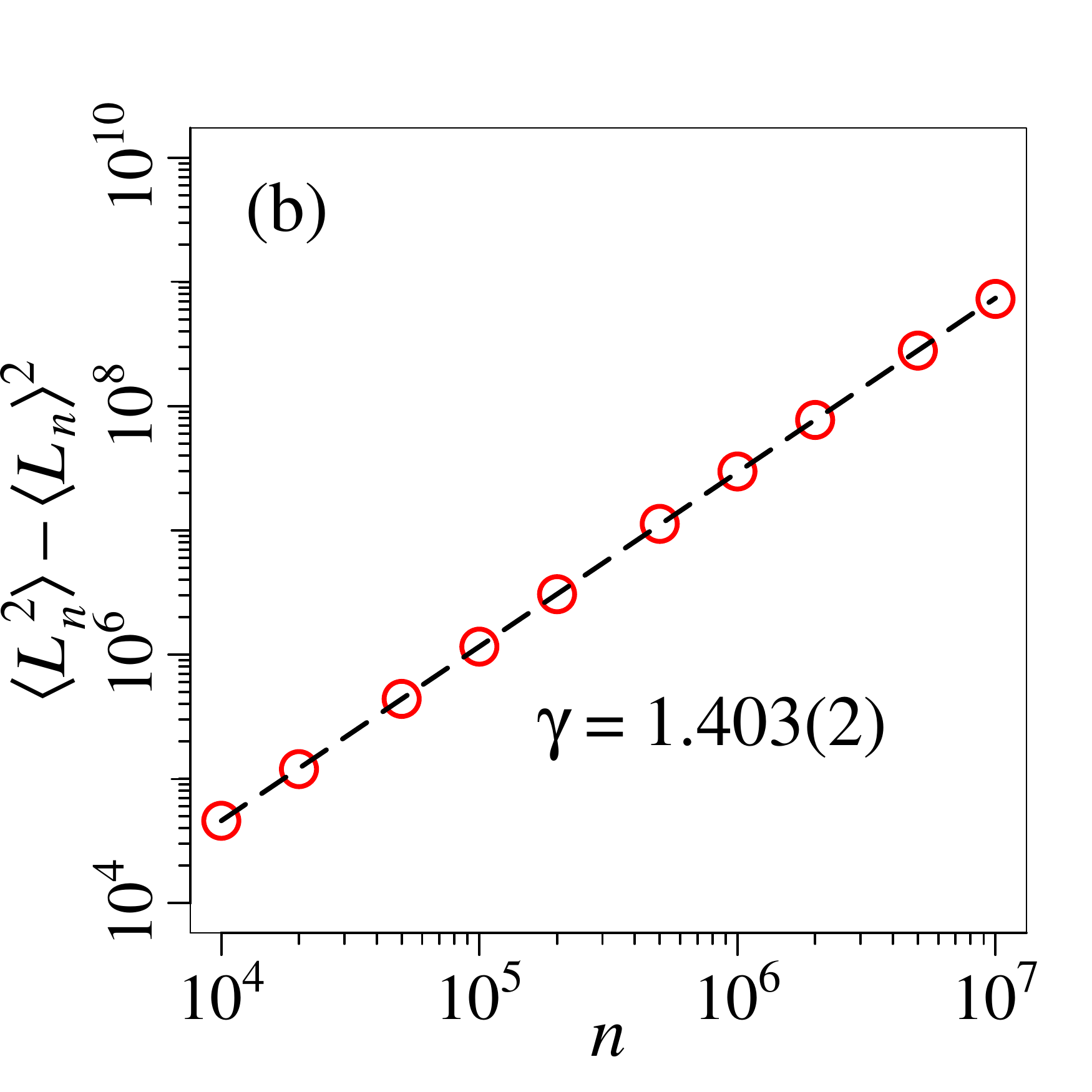}
\caption{Log-log plot of (a)~the empirical mean and (b)~the empirical variance of $L_{n}$ for the Student's $t$ random walk with parameter $\nu=2/3$ together with the least-squares fits (dashed lines). For this value of $\nu$, expression (\ref{eq:theta}) with $c=0$ yields the best fit. The fact that the curves have virtually the same slope in the different vertical scales on the graphs suggests that $\gamma \simeq 2\theta$. Each point was obtained from an average over $10^{4}$ sample random walks.}
\label{fig:loglog}
\end{figure}%
over the range of sizes investigated. Figure~\ref{fig:loglog} displays log-log plots of $\langle{L_{n}}\rangle$ and $\langle{L_{n}^{2}}\rangle-\langle{L_{n}}\rangle^{2}$ for $\nu=2/3$ for illustration. The least-squares fit estimates obtained for $\theta$ and $\gamma$ are shown in Table~\ref{tab:theta}. For $\nu<3/2$, the size of the logarithmic correction is very small, thus, we set it to zero to obtain the final value of $\theta$ (which reduced the error bars for $\theta$ here). Note that for $\nu\ge 2$ all values of $\theta$ obtained are roughly in the range $[0.49,0.51]$, i.\,e., ${\theta=1/2}$ with a logarithmic correction seems to hold, confirming previous predictions \cite{lisrw,hartmann}. In Section~\ref{sec:optimal}, however, we revisit the estimation of the constants appearing in (\ref{eq:conj}). For $\nu\le 3/2$, we obtained $\gamma \simeq 2\theta$ to a very good precision, in agreement with the picture provided by Fig.~\ref{fig:loglog}, suggesting that the probability density function (\pdf) of $L_{n}$ indeed follows the form (\ref{eq:scaling}), since then the $k$th moment of $L_{n}$ becomes
\begin{equation}
\label{eq:change}
\begin{split}
\mathbb{E}(L_{n}^{k}) = &\int\! L_{n}^{k} f(L_{n})\,dL_{n} = \\ &= \mathbb{E}(L_{n})^{k} \!\int z_{n}^{k} g(z_{n})\,dz_{n} = c_{n,k} \mathbb{E}(L_{n})^{k},
\end{split}
\end{equation}
with $z_{n}=L_{n}/\mathbb{E}(L_{n})$ and $c_{n,k}$ the $k$th moment of the distribution $g(z_{n})=\mathbb{E}(L_{n})f(\mathbb{E}(L_{n})z_{n})$. We see that, with $f(L_{n})$ like in Eq.~(\ref{eq:scaling}), all moments $\mathbb{E}(L_{n}^{k}) \propto \mathbb{E}(L_{n})^{k}$, as our data for $k=1$ and $2$ do for small enough values of $\nu$. Note that, as a consequence,
\begin{table}[b]
\renewcommand{\arraystretch}{0.7}
\caption{Exponents $\theta$ and $\gamma$ fitted according to (\ref{eq:theta}) and (\ref{eq:gamma}) for selected values of tail index $\nu$. The ratio $\gamma/\theta \simeq 2$ suggests form (\ref{eq:scaling}) [see also Eq.~(\ref{eq:change})] for the \pdf\ of $L_{n}$. The numbers between parentheses indicate the uncertainty in the last digit(s) of the data.}
\label{tab:theta}
\begin{ruledtabular}
\begin{tabular}{lccc}
$\nu$ & $\theta$ & $\gamma$ & $\gamma/\theta$ \\[4pt]
\colrule
1/2 &  $0.7050(6)$  &  $1.419(2)$   &  $2.01$  \\
1   &  $0.6850(6)$  &  $1.372(3)$   &  $2.00$  \\
3/2 &  $0.588(7)$   &  $1.355(13)$  &  $2.30$  \\
2   &  $0.516(3)$   &  $1.294(7)$   &  $2.51$  \\
5/2 &  $0.500(4)$   &  $1.259(9)$   &  $2.52$  \\
3   &  $0.496(3)$   &  $1.246(8)$   &  $2.51$  \\
7/2 &  $0.501(5)$   &  $1.260(16)$  &  $2.51$  \\
4   &  $0.504(4)$   &  $1.259(10)$  &  $2.40$  \\
5   &  $0.497(3)$   &  $1.257(10)$  &  $2.52$  \\
\end{tabular}
\end{ruledtabular}
\end{table}%
\begin{equation}
\label{eq:selfavg}
\var(L_{n}) = \mathbb{E}(L_{n}^{2})-\mathbb{E}(L_{n})^{2} = (c_{n,2}-c_{n,1}^{2})\,\mathbb{E}(L_{n})^{2},
\end{equation}
and the random variable $L_{n}$ cannot possibly be self-averaging unless $(c_{n,2}-c_{n,1}^{2}) \stackrel{n}{\longrightarrow} 0$, i.\,e., unless $g(z_{n})$ becomes increasingly more concentrated with $n$,
\begin{equation}
\label{eq:delta}
g(z_{n}) \stackrel{n}{\longrightarrow} \delta(z-c_{1}).
\end{equation}
Our data indicate, however, that $g(z_{n})$ remains broad irrespective of how large $n$ gets.

Figure~\ref{fig:theta} displays a log-log plot of $\theta$ against $\nu$. The plot does not suggest any clear functional relationship between $\theta$ and $\nu$---we were hoping for something like $\theta \sim (\nu_c-\nu)^{z}$ in the interval $\nu \leq 2$. Otherwise, $\theta$ saturates at $\theta \simeq 0.5$ for distributions of step lengths of finite variance ($\nu > 2$), with a transient behavior in the interval $2 < \nu \leq 5/2$ that we attribute to the finite length of the random walks ($n \leq 10^{7}$ steps).

Figure~\ref{fig:collapse} displays data collapses for our LIS data for some selected $\nu$ employing expressions (\ref{eq:exp}) or (\ref{eq:conj}), depending whether $\nu < 2$ or $\nu \geq 2$, respectively. The curve resulting from the data collapse corresponds to the empirical distribution $g(z)$ in Eq.~(\ref{eq:scaling}). We see very good data collapses, all virtually with the same form for $g(z_{n})$. We also see that it is definitely not the case that $g(z_{n}) \stackrel{n}{\longrightarrow} \delta(z)$ [cf.~Eqs.~(\ref{eq:change})--({\ref{eq:delta})].

\begin{figure}[t]
\centering
\includegraphics[viewport=20 10 480 450, scale=0.375, clip]{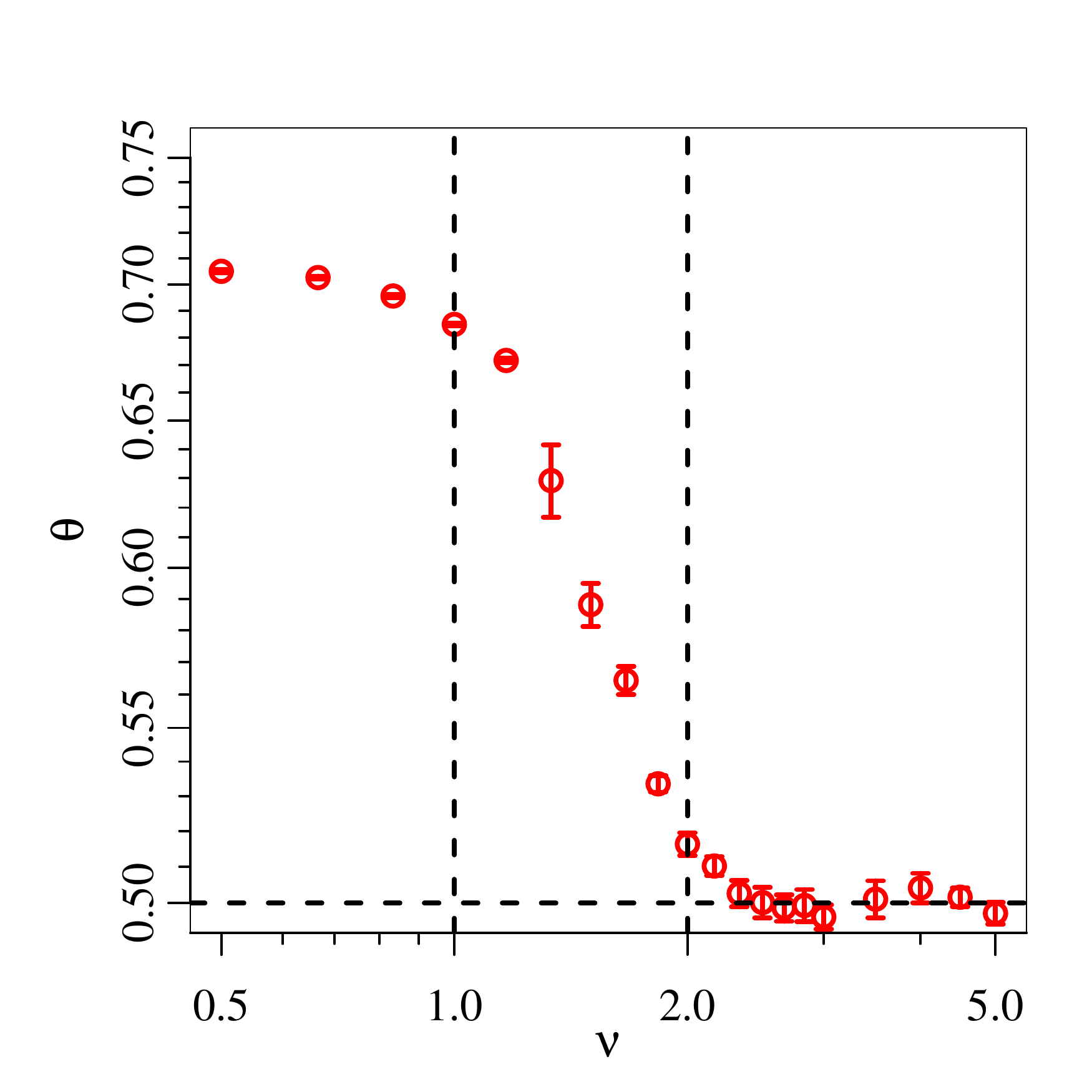}
\caption{Log-log plot of the scaling exponent $\theta$ against the tail index $\nu$. The dashed vertical lines delimit the intervals in which $\phi_{\nu}(\xi)$ [Eq.~(\ref{eq:student})] does not have finite integer moments ($\nu \leq 1$), has only finite mean ($1 < \nu \leq 2$), and has finite mean and variance ($\nu > 2$). The dashed horizontal line marks the $\theta=1/2$ line. For some data points the error bars are smaller than the symbols shown.}
\label{fig:theta}
\end{figure}

\begin{figure}[ht]
\centering
\includegraphics[viewport=0 10 480 460, scale=0.245, clip]{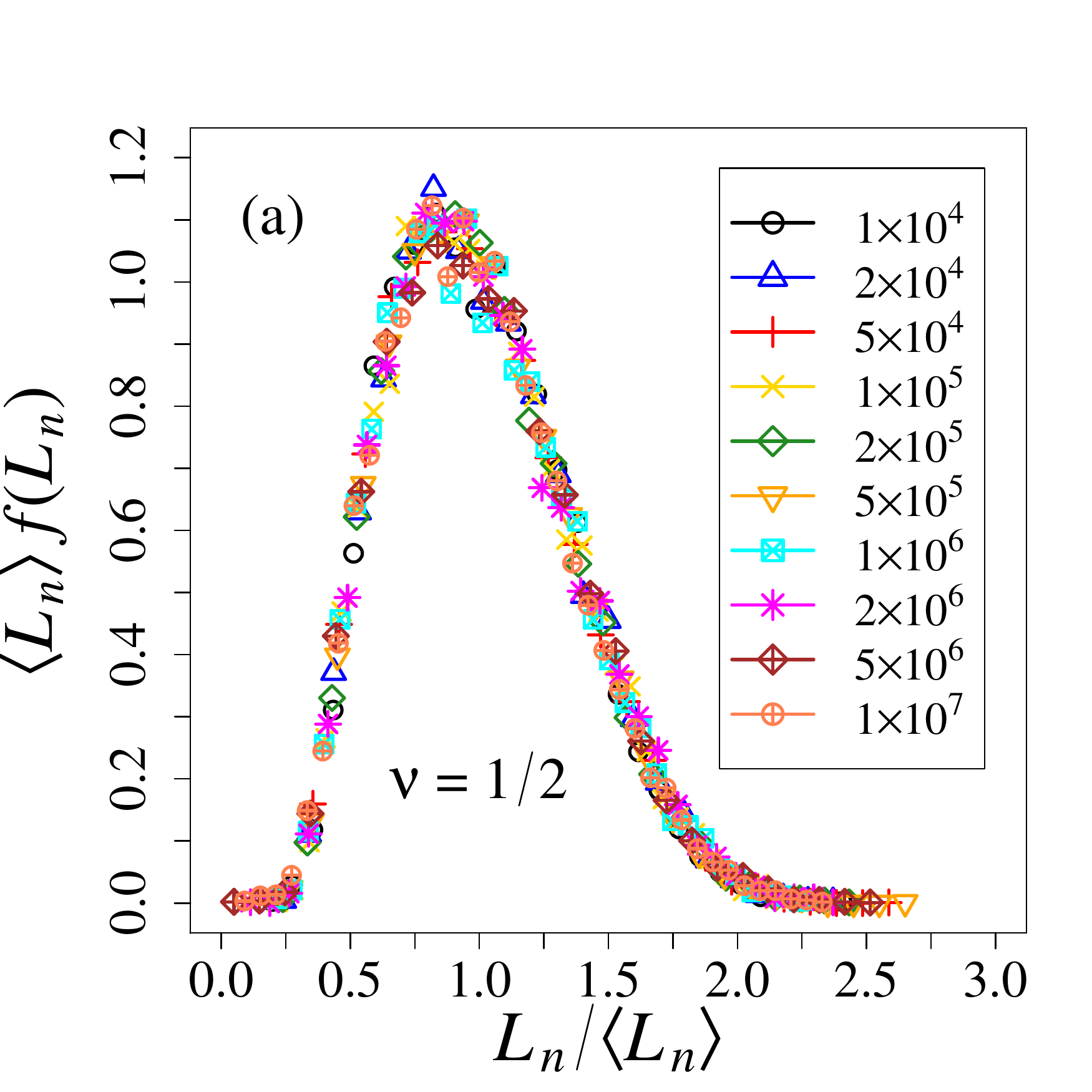}  \hfill
\includegraphics[viewport=0 10 480 460, scale=0.245, clip]{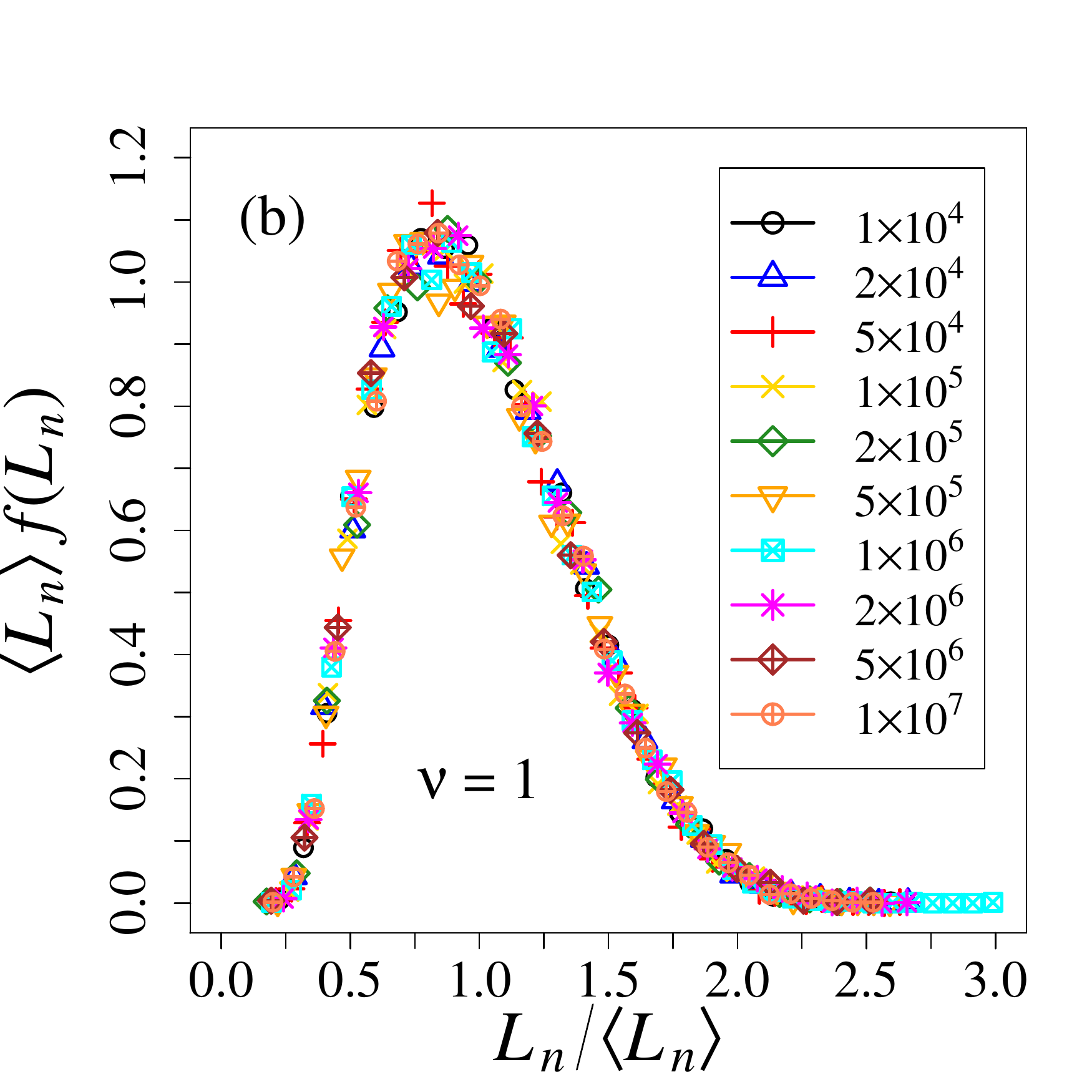} \includegraphics[viewport=0 10 480 460, scale=0.245, clip]{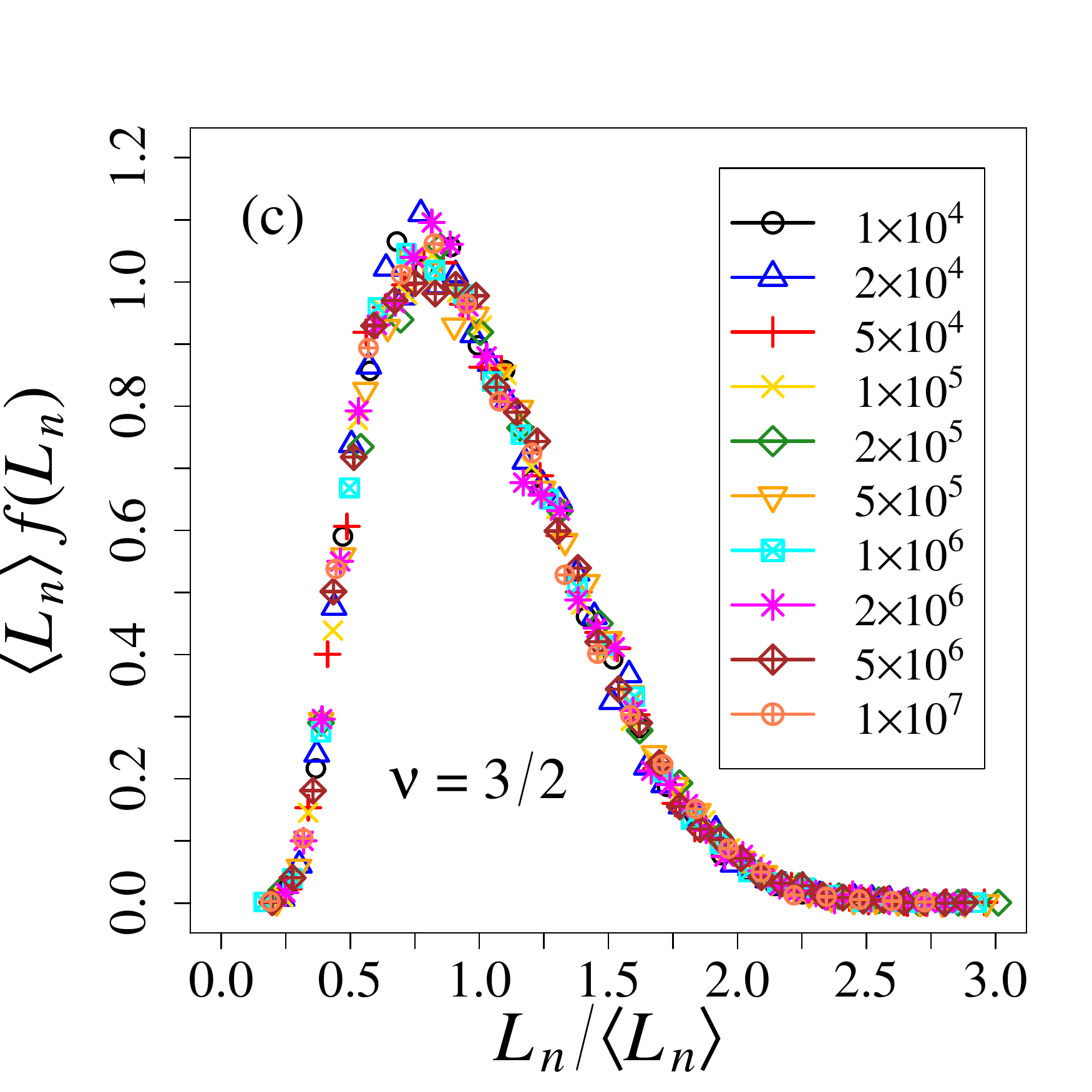}  \hfill
\includegraphics[viewport=0 10 480 460, scale=0.245, clip]{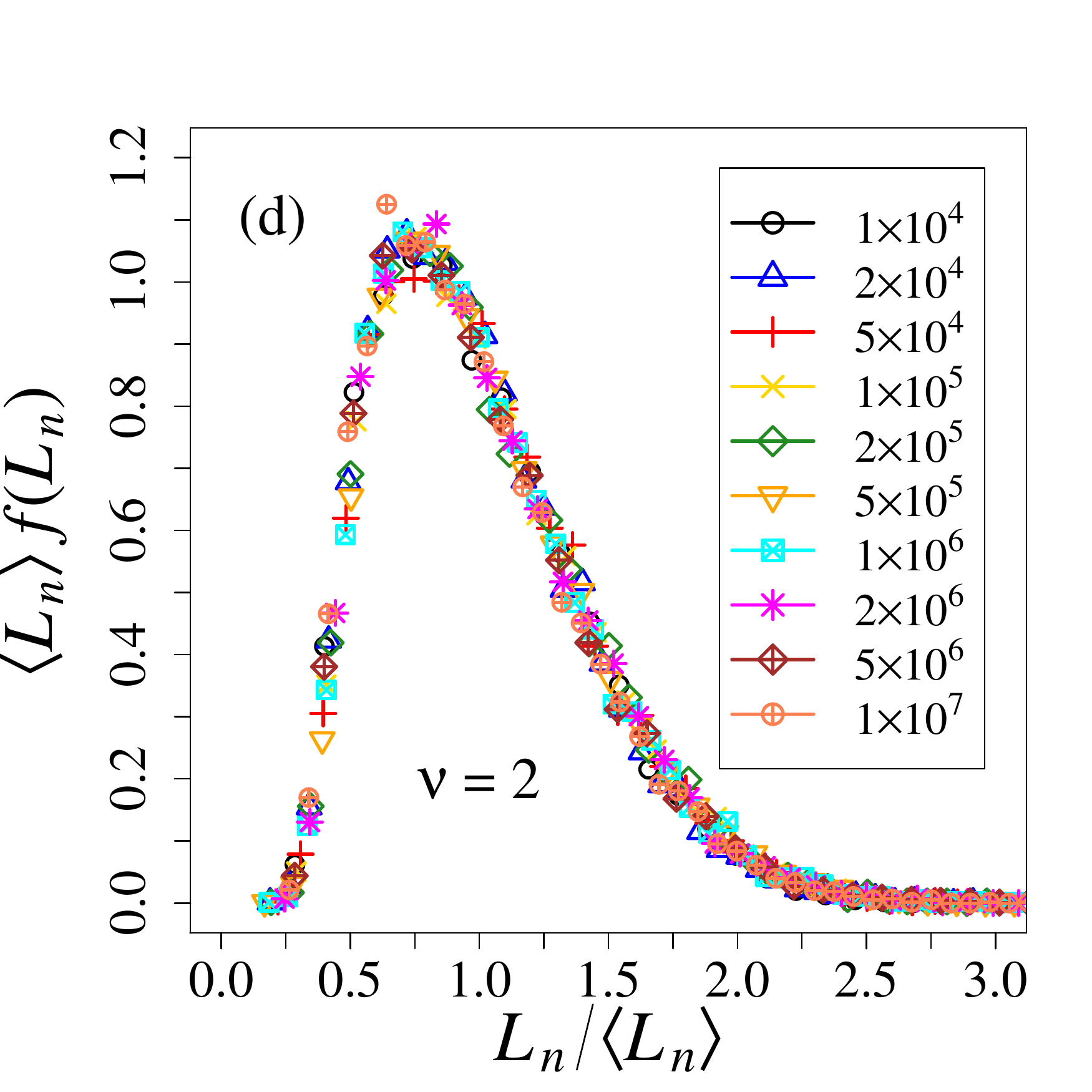} \includegraphics[viewport=0 10 480 460, scale=0.245, clip]{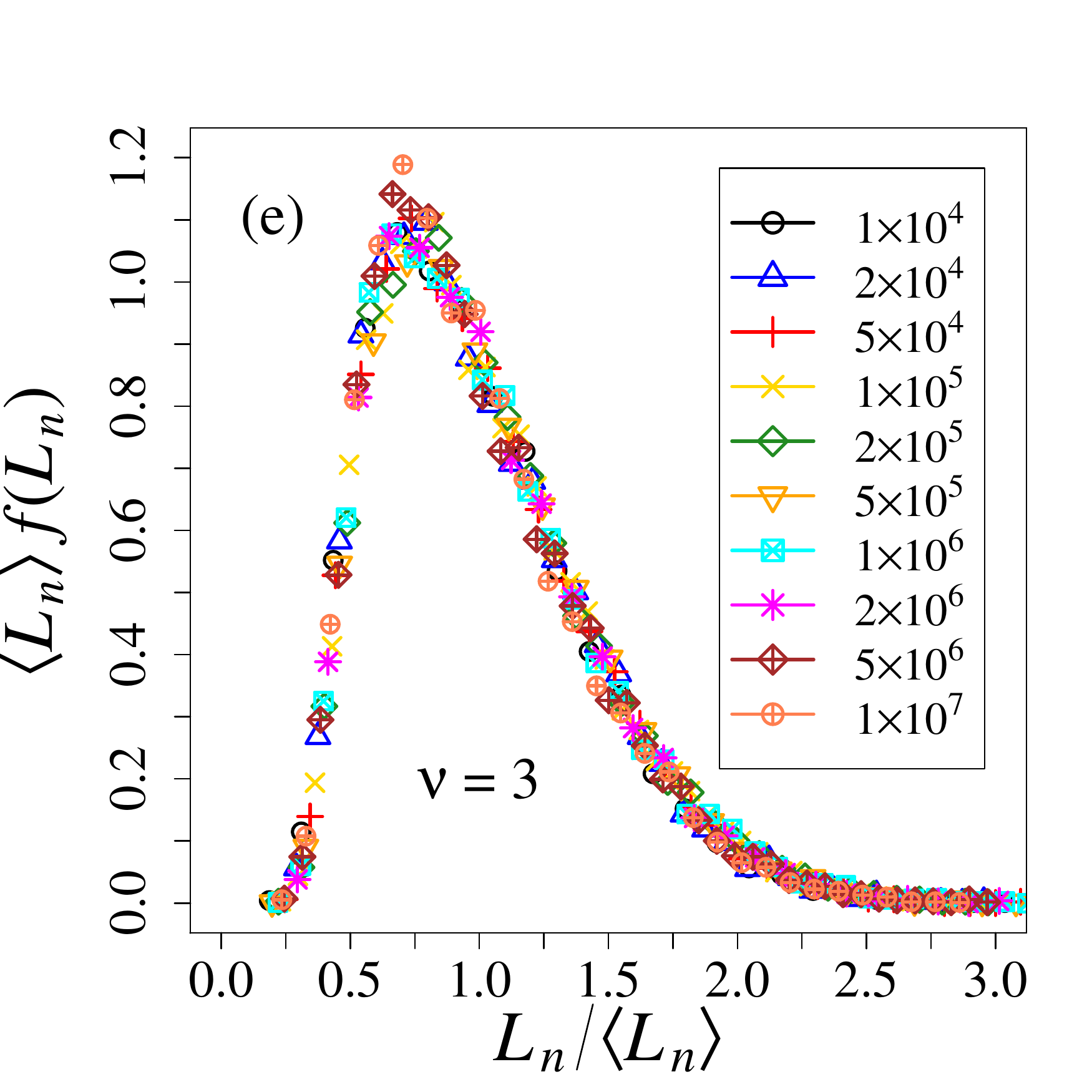} \hfill
\includegraphics[viewport=0 10 480 460, scale=0.245, clip]{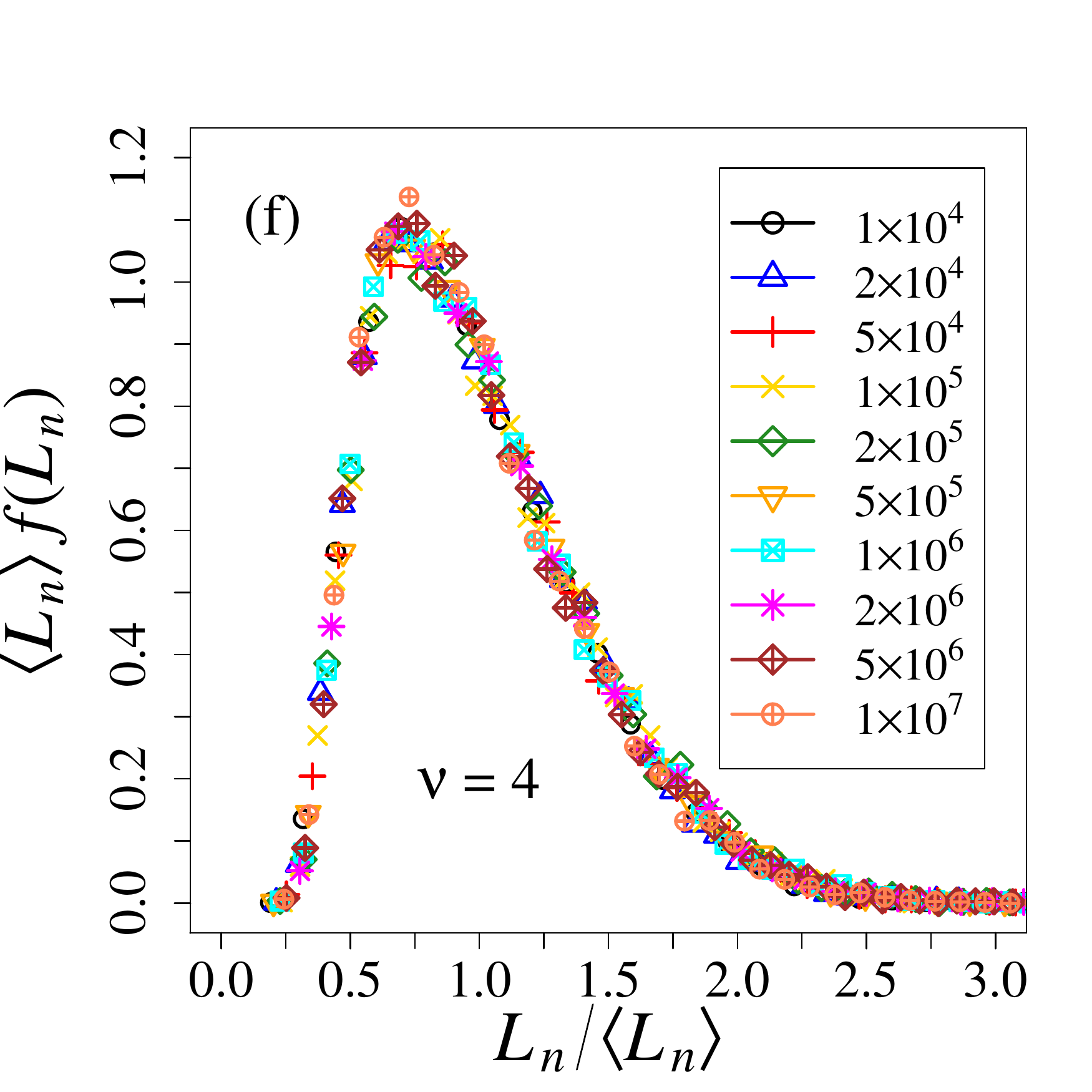}
\caption{Data collapses for some LIS data according to expressions (\ref{eq:exp}) and (\ref{eq:conj}). The upper panels (a) and (b) display data for the LIS of random walks with step lengths with both infinite mean and variance ($\nu \leq 1$), the middle panels (c) and (d) display data when the step increments have finite mean but infinite variance ($1 < \nu \leq 2$), and the bottom panels (e) and (f) display data obtained from random walks with step increments with both mean and variance finite ($\nu>2$).}
\label{fig:collapse}
\end{figure}


\section{\label{sec:remarks}\!The LIS of random walks of finite variance}

\subsection{\label{sec:wishful}A wishful (but unlikely) connection with integer partitions}

Recall that a partition of a natural number $n$ is a sequence of integers $\lambda_{1} \geq \cdots \geq \lambda_{k} > 0$ such that $\lambda_{1} + \cdots + \lambda_{k} = n$. For example, $(5,4,3)$ and $(4,4,3,1)$ are two partitions of $n=12$. No closed-form expression for the number $p(n)$ of partitions of $n$ is known. Asymptotically, for $n \to \infty$ we have the Hardy-Ramanujan formula $p(n) \sim \exp(\pi\sqrt{2n/3})/4\sqrt{3}n$ \cite{hardy}.

As is well known, integer partitions play an important role in the solution of the LIS problem for random permutations \cite{bdj99,patience,romik}. In this case, the partitions carry the Plancherel measure given by $\mathbb{P}(\lambda) = (\dim\lambda)^{2}/n!$, where $\dim\lambda$ is the dimension of the irreducible representation of the symmetric group $\mathfrak{S}_n$ indexed by $\lambda$ or, equivalently, the number of Young tableaux of shape $\lambda$. The correspondence between permutations and integer partitions (via the Robinson-Schensted-Knuth correspondence between permutations and Young tableaux) then allows one to identify the largest part of the partition $\lambda$ with the length of the LIS of the original permutation.

Clearly, other probability measures for random integer partitions have also been considered. Of particular interest to us is the uniform measure given by $\mathbb{P}(\lambda) = 1/p(n)$. This is so because the expected size of the largest part $\lambda_{1}$ of a partition of a large integer $n$ drawn from the set of all partitions of $n$ uniformly at random is given asymptotically by \cite{husimi,lehner,fristedt,vershik,bosegas,grabner}
\begin{equation}
\label{eq:part}
\mathbb{E}(\lambda_{1}) = \sqrt{\frac{n}{4\zeta(2)}}[\ln{n}+2\gamma_{E}-\ln{\zeta(2)}] + O(\ln{n}),
\end{equation}
where $\zeta(2) = \pi^{2}/6$ and $\gamma_{E} = 0.577215{\dots}$ is Euler's constant. Equation (\ref{eq:part}) has, to leading and subleading order, the same functional form as the conjectured expression (\ref{eq:conj}) for the asymptotic behavior of the length of the LIS of random walks with step lengths of finite variance. Moreover, the constant $1/\sqrt{4\zeta(2)} = 0.389848{\dots}$ accompanying the leading term of Eq.~(\ref{eq:part}) is close to the conjectured $1/e = 0.367879{\dots}$ in (\ref{eq:conj}). In the random partition model, however, the largest part $\lambda_{1}$ fluctuates, asymptotically for large $n$, like a Gumbel random variable with distribution
\begin{equation}
\label{eq:gumbel}
\mathbb{P}(\lambda_{1} \leq \lambda) = F_{G}\bigg(\frac{\lambda-\sqrt{n/4\zeta(2)}\ln\,[n/4\zeta(2)]}{2\sqrt{n/4\zeta(2)}}\bigg),
\end{equation}
where $F_{G}(z) = \exp[-\exp(-z)]$ \cite{lehner,fristedt,vershik,bosegas}. We can thus check whether Eq.~(\ref{eq:part}) makes sense in the context of the LIS problem for random walks beyond mere coincidence by checking whether our LIS data follow a Gumbel distribution.

\begin{figure}[t]
\centering
\includegraphics[viewport=20 10 480 450, scale=0.375, clip]{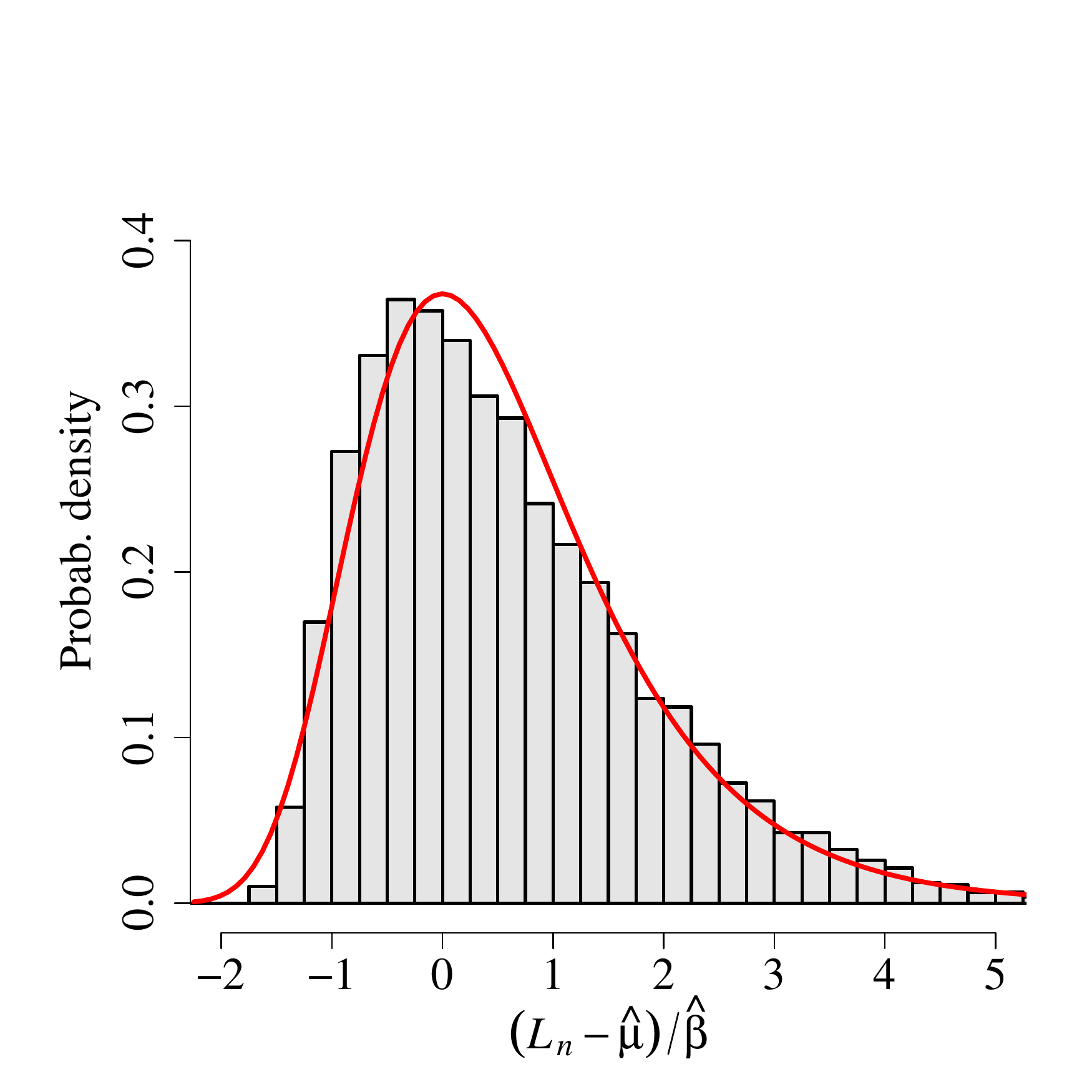}
\caption{Histogram of $L_{n}$ obtained from $10^{4}$ Gaussian random walks of $10^{8}$ steps each together with the standard Gumbel \pdf\ (solid line). The LIS data is centered and normalized by the adjusted parameters $\hat\mu$ and $\hat\beta$, respectively [see Eqs.~(\ref{eq:fxmb})--(\ref{eq:mubeta})].}
\label{fig:gumbel}
\end{figure}

The mean and variance of a Gumbel random variable $X$ with \pdf\
\begin{equation}
\label{eq:fxmb}
f_{G}(x;\mu,\beta) = \frac{1}{\beta}
\exp\Big[-\Big(\frac{x-\mu}{\beta}\Big) - \exp\Big(-\frac{x-\mu}{\beta}\Big)\Big]
\end{equation}
are given by
\begin{equation}
\mathbb{E}(X)=\mu+\gamma_{E}\beta, \quad \var(X) = \frac{\pi^{2}}{6}\beta^{2}.
\end{equation}
If we substitute the sample mean $\langle L_{n} \rangle = 69946$ for $\mathbb{E}(X)$ and the sample variance $\langle L_{n}^2 \rangle-\langle L_{n} \rangle^{2} \simeq 8.399 \times 10^{8}$ for $\var(X)$ of a LIS dataset obtained from $10^{4}$ Gaussian random walks of $n=10^{8}$ steps each, we obtain the following simple estimation for the parameters in (\ref{eq:fxmb}),
\begin{equation}
\label{eq:mubeta}
\hat\mu \simeq 56903, \quad \hat\beta \simeq 22597.
\end{equation}
We do not care about the uncertainties in $\hat\mu$ or $\hat\beta$ because they are relatively small and because the estimation procedure itself (the method of moments) is only approximate. While $\hat\mu$ (as well as $\langle L_{n} \rangle$) is not very far from the respective factor in Eq.~(\ref{eq:gumbel}) with $n=10^{8}$, to wit, $\mu = \sqrt{n/4\zeta(2)} \ln[n/4\zeta(2)] = 64468$, the value of $\hat\beta$ differs significantly from $\beta = 2\sqrt{n/4\zeta(2)} = 7797$. Figure~\ref{fig:gumbel} displays the histogram of the LIS data together with a plot of $f_{G}(z) = F_{G}'(z) = \exp[-z-\exp(-z)]$. The fit looks good, but not excellent. In fact, the Gumbel distribution has a skewness of $12\sqrt{6}\,\zeta(3)/\pi^3 = 1.139{\dots}$, while the data distribution has skewness $\simeq 0.976$ (irrespective of linear scaling). Whether this discrepancy is a finite-size effect is not clear at this moment. It should be remarked, however, that LIS data obtained from a uniform $U(-1,1)$ distribution of step increments with other values of walk length $n$ provide the same overall picture as in Fig.~\ref{fig:gumbel} and seems to be nearly independent of $n$.

We leave the quantification of the ``Gumbel hypothesis'' to a future study employing more sophisticated density estimation techniques and hypothesis testing to tame uncontrolled wishful thinking \cite{magical}. In Section~\ref{sec:optimal}, however, we will discover that Eq.~(\ref{eq:part}) cannot be easily discarded as a possible scaling form for the length of the LIS of random walks with step lengths of finite variance.


\subsection{\label{sec:optimal}Constraining the range of the parameters \\ in the asymptotic formula for $\mathbb{E}(L_{n})$}

The constants $1/e$ and $1/2$ appearing in the conjectured asymptotic formula (\ref{eq:conj}) for the $\mathbb{E}(L_{n})$ of random walks with distribution of step increments of finite variance were originally guessed in Ref.~\cite{lisrw} based on linear least-squares fits to the data. Equation~(\ref{eq:part}), with the same functional form, displays, respectively, the constants $1/\sqrt{4\zeta(2)} = 0.389848{\dots}$ and $[2\gamma_{E}-\ln{\zeta(2)}]/\sqrt{4\zeta(2)} = 0.256025{\dots}$. The proximity between the two sets of constants led us to reassess the numerical values of the constants appearing in (\ref{eq:conj}) by a more refined approach.

Assuming that Eq.~(\ref{eq:scaling}) holds for the distribution of $L_{n}$ and that $\mathbb{E}(L_{n})$ behaves asymptotically like [cf.~Eq.~(\ref{eq:conj}) or (\ref{eq:theta})]
\begin{align}
\label{eq:ab}
\mathbb{E}(L_{n}) \sim a\sqrt{n}\ln{n} + b\sqrt{n},
\end{align}
we can use the entire measured distributions to estimate the parameters $a$ and $b$ and give confidence intervals on their possible values. For a given pair $(a, b)$ we scale the data points according to Eq.~\eqref{eq:ab} and estimate the quality $S^{(c)}_{a,b}$ (superscript $(c)$ for collapse, discussed below) of the resulting data collapse by a method first introduced in \cite{kawashima} and refined in \cite{houdayer} in the context of the finite-size scaling analysis of phase transitions. The method works by estimating the best master curve on which the data points for different sizes $n$ should collapse. The quality $S^{(c)}_{a,b}$ is defined as the mean-square distance of the data points to the master curve in units of the standard error, similar to a $\chi^2$ test. If the data points are on average one standard error away from the estimated master curve, the data collapse will have a quality of $S^{(c)}_{a,b} = 1$. Values $S^{(c)}_{a,b} \ll 1$ indicate that the standard errors are overestimated; values $S^{(c)}_{a,b} \gg 1$ indicate that the data points do not collapse within error bars, i.\,e., that the quality of the data collapse is bad. A data collapse of bad quality might be due to, besides the inevitable errors in the estimation of the master curve, also finite-size effects in the data and corrections to the functional form (\ref{eq:ab}) itself.

In our case, care should be exercised in the application of the method because the scaling function $g(z)$ [see Eq.~(\ref{eq:scaling})] is insensitive to the multiplication of $L_{n}$, and thus $\mathbb{E}(L_{n})$, by a nonzero factor, i.\,e., to any rescaling $(a, b) \to (ra, rb)$ by some (real) $r \neq 0$. This leaves the determination of the optimal $(a, b)$ ill-defined. To fix this we compare the average value of each data set with the values predicted by Eq.~\eqref{eq:ab} for every pair $(a, b)$ tested. We then apply the same method as before to compute the quality figure $S^{(m)}_{a,b}$ (superscript $(m)$ for mean) using Eq.~\eqref{eq:ab} as the master curve with an added generous uncertainty of $\pm 0.05 \times \mathbb{E}(L_{n})$ to account for finite-size effects and possible lower-order terms. Note that the minimum of $S^{(m)}_{a,b}$ corresponds to a standard least-squares fit.

The above mentioned analyses were performed on data obtained from random walks of $n = 2^{16}$, $2^{17}$, $2^{18}$, and $2^{19}$ steps distributed according to a uniform $U(-1, 1)$ distribution; for each value of $n$, $10^6$ sample random walks (and thus $10^{6}$ data points $L_{n}$) are generated. Because data collapse is a matter of the form of a curve, we perform the collapse on $\ln f(L_{n})$. This means that instead of the absolute standard errors $\sigma$ of each data point we use the relative $\sigma/f(L_{n})$ instead.

\begin{figure}[t]
\centering
\includegraphics[scale=1.0]{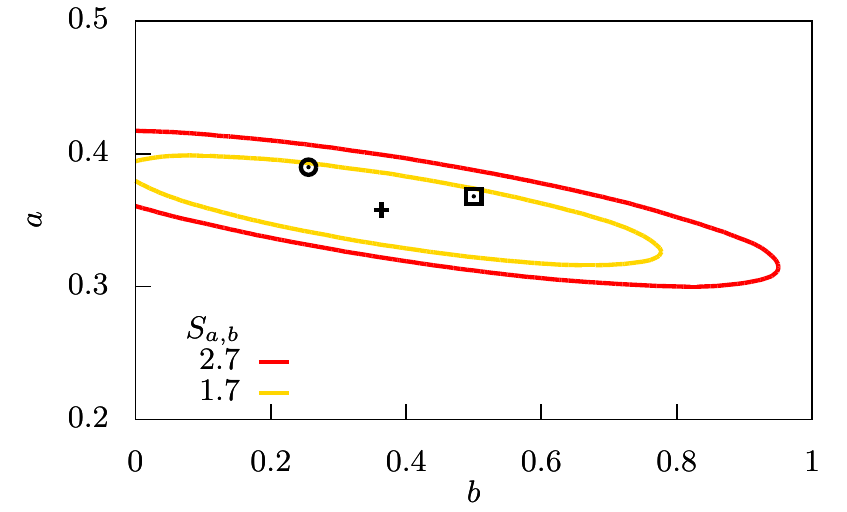}
\caption{Quality landscape $S_{a,b}$ obtained for random walks of $n = 2^{16}$, $2^{17}$, $2^{18}$, and $2^{19}$ uniformly, $U(-1, 1)$ distributed steps; $10^6$ random walks were generated for each $n$. The square symbol indicates the point $(1/e, 1/2)$ proposed in \cite{lisrw}, the circle indicates the point $\left(0.389848\cdots, 0.256025\cdots\right)$ corresponding to Eq.~(\ref{eq:part}), and the plus indicates the point $\left(0.36, 0.36 \right)$ at which $S_{a,b}$ attains its minimum.}
\label{fig:contours}
\end{figure}

Figure~\ref{fig:contours} displays the contour plot of the composite quality factor $S_{a,b} = \frac{1}{2}\big(S^{(c)}_{a,b} + S^{(m)}_{a,b}\big)$ for our data, which are collected by a scan through the $(a, b)$ space in discrete steps of $\Delta a = 0.003$ and $\Delta b = 0.01$ of the aforementioned procedure. The best quality $S^{(\mathrm{min})}_{a,b}\approx0.7$ was achieved for $(a, b) = (0.36, 0.36)$. The equiquality lines at $S_{a,b}^{(\mathrm{min})} + 1$ and $S_{a,b}^{(\mathrm{min})}+2$ can be roughly understood as $1\sigma$ and $2\sigma$ confidence intervals around the best quality $S_{a,b}^{(\mathrm{min})}$ \cite{kawashima,houdayer}. Note that the precise shape of the confidence intervals depends on the weighting of the terms in the composite quality; for instance, a larger weight on $S^{(m)}_{a,b}$, say, $\hat S_{a,b} = \frac{1}{4}\big(S^{(c)}_{a,b} + 3S^{(m)}_{a,b}\big)$, leads to a further elongation of the equiquality loci. Figure~\ref{fig:abcollapse} shows the data collapse as well as the least-squares fit which, in combination, yield the best quality. A further data collapse using high precision data from Ref.~\cite{hartmann} in the far tails of the distribution shows a different picture, with $S_{a,b}^{(\mathrm{min})}=81$ at $(a, b) = (0.33, 0.68)$. This data collapse, although still including the pair $(1/e, 1/2)$ within the second interval, is clearly of bad quality ($S_{a,b}^{(\mathrm{min})} \gg 1$), probably because of strong finite-size effects, since we only have data in the far tails of the distributions for small walk lengths $n \leq 4096$. This data collapse is therefore not shown.

\begin{figure}[t]
\centering
\includegraphics[scale=1.0]{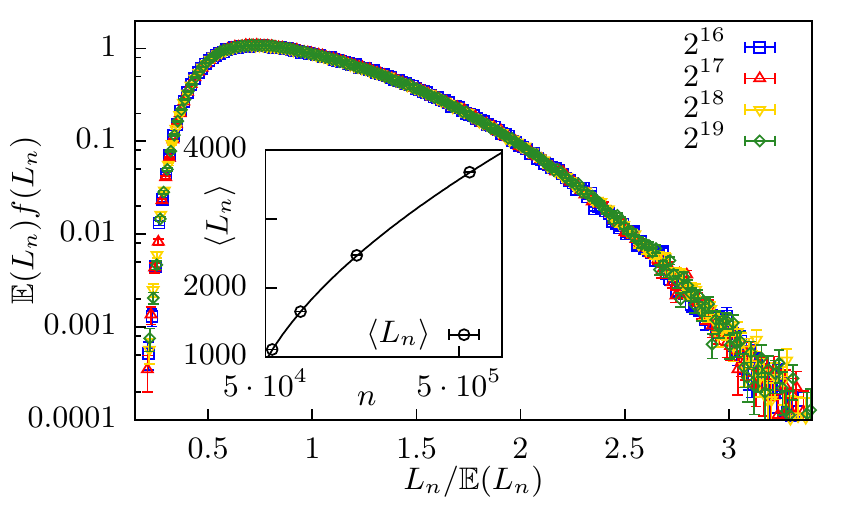}
\caption{Data collapse of the distributions with the optimal pair $(a, b) = (0.36, 0.36)$ that minimizes $S_{a,b}$. The inset shows the empirical averages $\langle L_{n} \rangle$ and their prediction according to Eq.~(\ref{eq:ab}) with the optimal constants $(a,b)=(0.36,0.36)$.}
\label{fig:abcollapse}
\end{figure}

We see from Fig.~\ref{fig:contours} that both pairs $(a, b) = (1/e, 1/2)$, which was proposed in \cite{lisrw} based on least-squares fits of mean values $\langle L_{n} \rangle$ for the LIS of Gaussian random walks, and $(a, b) = (0.389848\cdots, 0.256025\cdots)$ of Eq.~(\ref{eq:part}) lie within the second confidence interval suggested by our method. Since the data sources are independent and even originate from different distributions of step increments, we see this as an argument in favor of the proposed scaling form (\ref{eq:ab}).


\section{\label{sec:summary}Summary and conclusions}

We have extended previous studies on the length $L_{n}$ of the LIS of heavy-tailed random walks by considering Student's $t$-distributions with several different values of the parameter $\nu$. We found that $L_{n}$ scales like $\mathbb{E}(L_{n}) \sim n^{\theta}$ with a non-universal $\theta$ when $\phi(\xi)$ has infinite variance, but could not find a clear relationship $\theta = \theta(\nu)$ between these quantities besides the decreasing behavior $\theta'(\nu)<0$. We were expecting to find something like $\theta(\nu) \sim (\nu-\nu_{c})^{z}$ for some $\nu_{c}$, possibly equal to $1$ or $2$, and some ``critical exponent'' $z$. Unfortunately, though, as Fig.~\ref{fig:theta} indicates (note the log-log scale), such a relationship would entail a varying exponent $z$. Whether this is a finite-$n$ artifact we cannot tell. To seriously consider the hypothesis of a continuous phase transition for the behavior of the LIS between the $\nu > \nu_{c}$ and $\nu < \nu_{c}$ regions with $\theta$ as the order parameter one would probably have to simulate much longer random walks and possibly also larger sample sizes. Since we do not observe any strong change in the values of $\theta$ when increasing the sequence $n$ by a factor of 10, the walks probably would have to be by a factor of many decades longer, much beyond the available computational ressources.

When $\phi(\xi)$ is of finite variance ($\nu > 2$), we recover the asymptotic behavior given by (\ref{eq:conj}), but with newly estimated constants. Our best current estimates for the constants appearing in expression (\ref{eq:ab}), based on a sophisticated combined consideration of the behavior of the mean and of the scaling of the full distribution, are $a=0.36(3)$ and $b=0.36(30)$.

We could not obtain data for $\nu < 1/2$. The simulation of very-heavy-tailed random walks is complicated by the fact that one needs to add numbers of widely different orders of magnitude while keeping their full significance. This can be done with numerical libraries that implement arbitrary precision arithmetic, but the efficiency of the simulations suffers enormously. It would be desirable to compute the LIS of heavy-tailed random walks in the $\nu \to 0$ limit to check how $\theta$ scales with $\nu$ in this limit and how it compares with the bounds in (\ref{eq:heavy}).

Our results about the scaling behavior of the mean LIS length, although useful on their own, can find applications, for instance, in the descriptive analysis of time series: if the LIS length grows stronger than expected, this is an indication that hidden trends or correlations are present in the data.

While the Plancherel distribution of the largest part of an integer partition coincides with the distribution of the length of the LIS of a uniformly distributed random permutation \cite{bdj99,patience,romik}, the similarity between (\ref{eq:conj}) and Eq.~(\ref{eq:part}) does not imply any obvious relationship between the LIS of random walks and random integer partitions under the uniform measure. Our best estimated constants $a$ and $b$ for expression (\ref{eq:ab}), however, cannot rule out Eq.~(\ref{eq:part}) as a good candidate scaling form for the length of the LIS of random walks with step lengths of finite variance, and whether the LIS of these random walks follows a Gumbel distribution is open to debate. In a further study we intend to apply more refined density estimation techniques in the selection of an empirical model for the data; knowledge of the tail behavior, as provided by \cite{hartmann}, is a valuable piece of information in this regard.

The elucidation of a possible combinatorial structure behind the LIS of random walks remains a tantalizing issue.


\section*{Acknowledgments}

The authors thank Alexey Bufetov (U.~Bonn, Germany) for having called their attention to random integer partitions under the uniform measure, Pavel L. Krapivsky (Boston U., USA) and Satya N. Majumdar (LPTMS, France) for useful conversations, and two anonymous referees for several suggestions improving the manuscript. They also acknowledge usage of the HPC facilities of the GWDG G\"{o}ttingen and the CARL cluster in Oldenburg funded by the DFG grant no.~INST 184/157-1 FUGG and the Ministry of Science and Culture (MWK) of the Lower Saxony State. J.\,R.\,G.\,M. thanks the LPTMS for kind hospitality during a sabbatical leave in France and FAPESP (Brazil) for partial support through grant no.~2017/22166-9. H.\,S. and A.\,K.\,H. acknowledge support by DFG grant no.~HA~3169/8-1.



\begin{thebibliography}{00}

\bibitem{sergei}
S. Bespamyatnikh and M. Segal, Enumerating longest increasing subsequences and patience sorting, {Inf. Process. Lett.} \textbf{76}~(1--2), 7--11 (2000).

\bibitem{nevzorov}
V. B. Nevzorov, \textit{Records: Mathematical Theory} (AMS, Providence, 2001).

\bibitem{godreche}
C. Godr\`{e}che, S. N. Majumdar, and G. Schehr, Record statistics of a strongly correlated time series: random walks and L\'{e}vy flights, {J. Phys. A: Math. Theor.} \textbf{50}~(33), 333001 (2017).

\bibitem{gopalan2007}
P. Gopalan, T. Jayram, R. Krauthgamer, and R. Kumar, Estimating the sortedness of a data stream, in \textit{SODA~'07: Proceedings of the Eighteenth Annual ACM-SIAM Symposium on Discrete Algorithms}, New Orleans, LA, 2007 (SIAM, Philadelphia, 2007), pp.~318--327.

\bibitem{bonomi2016}
L. Bonomi and L. Xiong, On differentially private longest increasing subsequence computation in data stream, {Trans. Data Priv.} \textbf{9}~(1), 73--100 (2016).

\bibitem{ulam}
S. M. Ulam, Monte Carlo calculations in problems of mathematical physics, in \textit{Modern Mathematics for the Engineer: Second Series}, edited by E. F. Beckenbach (McGraw-Hill, New York, 1961), pp.~261--281.

\bibitem{t-widom}
C. A. Tracy and H. Widom, Level-spacing distributions and the Airy kernel, {Commun. Math. Phys.} \textbf{159}~(1), 151--174 (1994).

\bibitem{bdj99}
J. Baik, P. Deift, and K. Johansson, On the distribution of the length of the longest increasing subsequence of random permutations, {J. Am. Math. Soc.} \textbf{12}~(4), 1119--1178 (1999).

\bibitem{patience}
D. Aldous and P. Diaconis, Longest increasing subsequences: from patience sorting to the Baik-Deift-Johansson theorem, {Bull. Am. Math. Soc.} \textbf{36}~(4), 413--432 (1999).

\bibitem{romik}
D. Romik, \textit{The Surprising Mathematics of Longest Increasing Subsequences} (Cambridge University Press, New York, 2015).

\bibitem{angel}
O. Angel, R. Balka, and Y. Peres, Increasing subsequences of random walks, {Math. Proc. Cambridge} \textbf{163}~(1), 173--185 (2017).

\bibitem{pemantle}
R. Pemantle and Y. Peres, Non-universality for longest increasing subsequence of a random walk, {ALEA Lat. Am. J. Probab. Math. Stat.} \textbf{14}, 327--336 (2017).

\bibitem{newman1999}
M. E. J. Newman and G. T. Barkema, \textit{Monte Carlo Methods in Statistical Physics} (Clarendon, Oxford, 1999).

\bibitem{bigguide}
A. K. Hartmann, \textit{Big Practical Guide to Computer Simulations} (World Scientific, Singapore, 2015).

\bibitem{lisrw}
J. R. G. Mendon\c{c}a, Empirical scaling of the length of the longest increasing subsequences of random walks, {J. Phys. A: Math. Theor.} \textbf{50}~(8), 08LT02 (2017).

\bibitem{hartmann}
J. B\"{o}rjes, H. Schawe, and A. K. Hartmann, Large deviations of the length of the longest increasing subsequence of random permutations and random walks, {Phys. Rev. E} \textbf{99}~(4), 042104 (2019).

\bibitem{satya}
S. N. Majumdar and G. Schehr, Top eigenvalue of a random matrix: large deviations and third order phase transition, {J. Stat. Mech.: Theory Exp.} \textbf{(2014)}, P01012 (2014).

\bibitem{student}
Wikipedia contributors, \href{https://en.wikipedia.org/wiki/Student's_t-distribution}{Student's $t$-distribution} --- Wikipedia, The Free Encyclopedia, accessed 15 April 2019.

\bibitem{bailey}
R. W. Bailey, Polar generation of random variates with the $t$-distribution, {Math. Comput.} \textbf{62}~(206), 779--781 (1994).

\bibitem{hardy}
G. H. Hardy and S. Ramanujan, Asymptotic formul\ae\ in combinatory analysis, {Proc. Lond. Math. Soc. (2)} \textbf{17}, 75--115 (1918).

\bibitem{husimi}
K. Husimi, Partitio numerorum as occurring in a problem of nuclear physics, {Proc. Phys.-Math. Soc. Japan (3rd Ser.)} \textbf{20}, 912--925 (1938).

\bibitem{lehner}
P. Erd\H{o}s and J. Lehner, The distribution of the number of summands in the partitions of a positive integer, {Duke Math. J.} \textbf{8}~(2), 335--345 (1941).

\bibitem{fristedt}
B. Fristedt, The structure of random partitions of large integers, {Trans. Am. Math. Soc.} \textbf{337}~(2), 703--735 (1993).

\bibitem{vershik}
A. M. Vershik and Yu. Yakubovich, Fluctuations of the maximal particle energy of the quantum ideal gas and random partitions, {Commun. Math. Phys.} \textbf{261}~(3), 759--769 (2006).

\bibitem{bosegas}
A. Comtet, P. Leboeuf, and S. N. Majumdar, Level density of a Bose gas and extreme value statistics, {Phys. Rev. Lett.} \textbf{98}~(7), 070404 (2007); A. Comtet, S. N. Majumdar, and S. Ouvry, Integer partitions and exclusion statistics, {J. Phys. A: Math. Theor.} \textbf{40}~(37), 11255--11269 (2007).

\bibitem{grabner}
P. J. Grabner, A. Knopfmacher, and S. Wagner, A general asymptotic scheme for the analysis of partition statistics, {Combin. Probab. Comput.} \textbf{23}~(6), 1057--1086 (2014).

\bibitem{magical}
P. Diaconis, Theories of data analysis: From magical thinking through classical statistics, in \textit{Exploring Data Tables, Trends and Shapes}, edited by D. C. Hoaglin, F. Mosteller, and J. W. Tukey (Wiley, New York, 1985), pp.~1--36.

\bibitem{kawashima}
N. Kawashima and N. Ito, Critical behavior of the three-dimensional $\pm J$ model in a magnetic field, {J. Phys. Soc. Japan} \textbf{62}~(2), 435--438 (1993).

\bibitem{houdayer}
J. Houdayer and A. K. Hartmann, Low-temperature behavior of two-dimensional Gaussian Ising spin glasses, {Phys. Rev. B} \textbf{70}~(1), 014418 (2004).

\end{thebibliography}
\end{document}